\documentclass[10pt,twocolumn,letterpaper]{article}

\usepackage[pagenumbers]{wacv} 

\usepackage{wacv}

\newtoggle{wacvrebuttal}
\togglefalse{wacvrebuttal}

\usepackage[T1]{fontenc}
\usepackage[utf8]{inputenc}
\usepackage{multirow}

\definecolor{wacvblue}{rgb}{0.21,0.49,0.74}
\usepackage[pagebackref,breaklinks,colorlinks,allcolors=wacvblue]{hyperref}
\usepackage{booktabs}
\usepackage{xspace}
\usepackage{xcolor}
\usepackage{listings}

\newcommand{\methodname}{\textsc{PrivacyBench}\xspace}

\lstdefinelanguage{YAML}{
  basicstyle=\ttfamily\footnotesize,
  comment=[l]{\#},
  keywordstyle=\color{blue}\bfseries,
  stringstyle=\color{red},
  morekeywords={true,false,null}, 
  sensitive=false,
}

\definecolor{excellent}{RGB}{0,128,0}
\definecolor{good}{RGB}{0,100,0} 
\definecolor{poor}{RGB}{200,0,0}
\definecolor{bad}{RGB}{255,0,0}

\def\wacvPaperID{3154} 
\def\confName{WACV}
\def\confYear{2026}

\title{\methodname: Privacy Isn't Free in Hybrid Privacy-Preserving Vision Systems}

\author{
Nnaemeka Obiefuna$^{1,2\dagger}$ \,
Samuel Oyeneye$^{1*}$ \,
Similoluwa Odunaiya$^{1*}$ \,
Iremide Oyelaja$^{1*}$ \,
Steven Kolawole$^{1,3}$\\[4pt]
{\small
$^{1}$ML Collective \quad
$^{2}$Friedrich-Alexander-Universität Erlangen-Nürnberg \quad
$^{3}$Carnegie Mellon University
}\\[4pt]
nnaemeka.obiefuna@fau.de \quad
}

\begin{document}
\maketitle
\begingroup
\renewcommand\thefootnote{}\footnote{$\dagger$ First author}\addtocounter{footnote}{-1}
\endgroup

\begingroup
\renewcommand\thefootnote{}\footnote{* Equal contribution}\addtocounter{footnote}{-1}
\endgroup

\begin{abstract}
    Privacy-preserving machine learning deployments in sensitive deep learning applications—from medical imaging to autonomous systems—increasingly require combining multiple techniques. Yet, practitioners lack systematic guidance to assess the synergistic and non-additive interactions of these hybrid configurations, relying instead on isolated technique analysis that misses critical system-level interactions. 
    We introduce \methodname, a benchmarking framework that reveals striking failures in privacy technique combinations with severe deployment implications. Through systematic evaluation across ResNet18 and ViT models on medical datasets, we uncover that FL+DP combinations exhibit severe convergence failure---accuracy drops from 98\% to 13\% while compute costs and energy consumption substantially increased. In contrast, FL+SMPC maintains near-baseline performance with modest overhead. Our framework provides the first systematic platform for evaluating privacy-utility-cost trade-offs through automated YAML configuration, resource monitoring, and reproducible experimental protocols. 
    \methodname enables practitioners to identify problematic technique interactions before deployment, moving privacy-preserving computer vision from ad-hoc evaluation toward principled systems design. These findings demonstrate that privacy techniques cannot be composed arbitrarily and provide critical guidance for robust deployment in resource-constrained environments.
\end{abstract}    
\section{Introduction}

\begin{figure*}[t!]
\vspace{-5mm}
\centering
\includegraphics[width=1.0\textwidth]{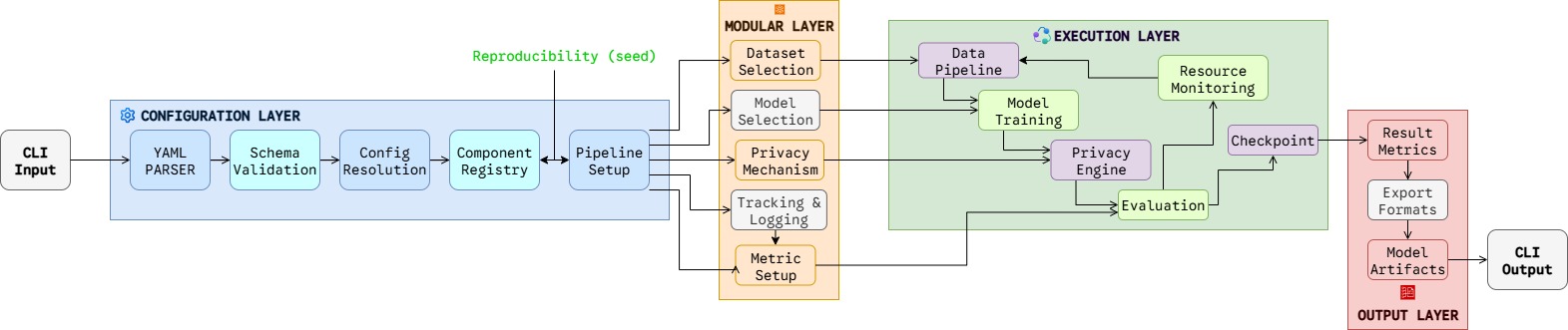}
\vspace{-4mm}
\caption{\textbf{\methodname Architecture Overview.} A four-layer modular framework enabling systematic evaluation of privacy technique interactions. The Configuration Layer handles YAML-based experiment specification, the Modular Layer supports diverse privacy combinations, the Execution Layer integrates comprehensive resource monitoring, and the Output Layer generates reproducible results. This architecture enables controlled evaluation of hybrid privacy configurations while tracking computational costs and energy consumption—capabilities missing from existing evaluation frameworks.}
\label{fig:architecture}
\vspace{-4mm}
\end{figure*}

Machine learning systems processing sensitive visual data---diagnostic models across hospitals, road scene analysis for autonomous vehicles, crowd monitoring for surveillance---increasingly require combining Privacy-Preserving Machine Learning (PPML) techniques to balance utility with protection requirements. Current deployments combine Federated Learning (FL) \cite{mcmahan2017communication}, Differential Privacy (DP) \cite{dwork2006differential, dwork2014algorithmic}, and Secure Multi-Party Computation (SMPC) \cite{cramer2015secure}, yet systematic evaluation frameworks for these technique combinations remain under-explored~\cite{liu2024survey,saha2024survey}, with practitioners lacking guidance for assessing their interactions in real-world vision systems.

Previous research approaches PPML from a purely algorithmic perspective, focusing on theoretical privacy guarantees and accuracy trade-offs while abstracting away practical system overheads—compute load, energy consumption, communication latency—that determine deployment feasibility~\cite{rajput2024enhancing,tschand2025mlperf}. The assumption that privacy technique costs are additive (e.g., FL overhead + DP overhead = total cost) represents a dangerous simplification. Complex interactions between techniques can produce non-linear effects on both model utility and system efficiency, with consequences that vary significantly across vision architectures~\cite{beyer2019reliable}.

This evaluation gap has severe deployment implications. Privacy technique combinations exhibit complex behaviors that individual assessments cannot predict: some combinations like FL+SMPC maintain near-baseline performance with modest overhead, while others like FL+DP can trigger severe performance degradation (98\% to 13\% accuracy) while increasing computational costs dramatically (20× energy consumption, 16× training time). Moreover, vision architectures exhibit distinct interaction patterns—transformer models show efficiency gains under federated training that CNNs do not demonstrate. Current evaluation practices assess privacy methods in isolation \cite{caldas2018leaf, wei2023dpmlbench, lai2022fedscale}, missing these critical interaction effects that determine production feasibility.

We introduce \methodname\footnote{\href{}{https://github.com/Federated-Learning-MLC/privacybench-exp}}, a systematic benchmarking framework (Fig.~\ref{fig:architecture}) that quantifies the full system cost of privacy-preserving vision deployments. Our framework addresses evaluation gaps through: systematic evaluation of privacy technique combinations using established frameworks (Flower \cite{beutel2022flower}, Opacus \cite{yousefpour2021opacus}); integrated resource monitoring including energy tracking via CodeCarbon \cite{codecarbon2021, lacoste2019quantifying}; comprehensive instrumentation of training time, memory usage, and convergence behavior; and YAML-based configuration management enabling standardized comparative studies across privacy techniques and model architectures. 
Our contributions are:
\begin{enumerate}
\item \textbf{\methodname framework}: Reproducible benchmarking platform with comprehensive resource monitoring and YAML-based configuration management for systematic evaluation of privacy technique combinations.
\item \textbf{Systematic evaluation methodology}: First comprehensive analysis of hybrid PPML configurations across vision architectures (ResNet18, ViT) and medical imaging datasets, measuring utility, computational cost, and energy footprint.
\item \textbf{Privacy technique interaction analysis}: Identification of both successful combinations (FL+SMPC) and critical failure modes (FL+DP), revealing architectural dependencies and resource consumption patterns that inform robust privacy system design.
\end{enumerate}

Through systematic evaluation on medical imaging datasets, we provide the first comprehensive analysis of privacy-utility-cost trade-offs in realistic deployment scenarios. Our findings reveal that privacy techniques exhibit predictable compatibility patterns based on operational abstractions—techniques operating at compatible levels compose successfully, while conflicting assumptions produce catastrophic failures. These insights enable informed deployment decisions and challenge assumptions about modular privacy system design.
\section{Related Work}
Prior efforts on privacy-preserving ML (PPML) and benchmarking fall into three silos. 
(1) \emph{Production PPML techniques} such as FL~\cite{mcmahan2017communication}, DP~\cite{dwork2006differential,abadi2016deep}, and SMPC~\cite{goldreich1987secure,shamir1979share,gentry2009fully} are widely deployed, but their hybrid combinations remain underexplored. 
(2) \emph{Benchmarking frameworks} like MLPerf~\cite{mattson2019mlperf,reddi2020mlperf,tschand2025mlperf} and vision benchmarks~\cite{deng2009imagenet,lin2014microsoft,beyer2019reliable,bangalore2024convision} standardize evaluation, yet none capture the system-level effects of privacy. 
(3) \emph{Privacy-specific frameworks} such as LEAF~\cite{caldas2018leaf}, FedScale~\cite{lai2022fedscale}, DPMLBench~\cite{wei2023dpmlbench}, and domain tools~\cite{zhou2024ppml,li2023privlm,apple2024pfl} evaluate single techniques in isolation, without considering energy costs~\cite{rajput2024enhancing, lacoste2019quantifying, codecarbon2021}. 
Together these silos leave a gap: no framework systematically evaluates hybrid privacy configurations with full resource monitoring. 
We review each strand in turn (\S\ref{lit:part1}--\ref{lit:part4}) before positioning \methodname as filling this gap (\S\ref{lit:part5}).

\subsection{PPML in Production Systems}
\label{lit:part1}
Modern privacy-sensitive applications increasingly combine multiple privacy techniques to meet diverse security requirements. \emph{Federated Learning} (FL), pioneered by McMahan et al.~\cite{mcmahan2017communication}, enables collaborative training across decentralized data owners without explicit sharing, with production implementations like Flower~\cite{beutel2022flower} supporting multi-party medical imaging and financial analytics. \emph{Differential Privacy} (DP), formalized by Dwork~\cite{dwork2006differential} and adapted for deep learning by Abadi et al.~\cite{abadi2016deep}, provides provable guarantees via noise injection, implemented in practice by libraries such as Opacus~\cite{yousefpour2021opacus}. \emph{Secure Multi-Party Computation} (SMPC), introduced by Goldreich et al.~\cite{goldreich1987secure}, protects data during computation using secret sharing~\cite{shamir1979share} and homomorphic encryption~\cite{gentry2009fully}.

Production deployments increasingly combine these techniques
such as FL+DP~\cite{geyer2017differentially,wei2020federated} or FL+SMPC~\cite{liu2024survey} to satisfy regulatory and adversarial requirements. Healthcare collaborations often require both decentralized training and formal guarantees, while financial systems rely on federated coordination with cryptographic protection. Yet surveys highlight a persistent gap: while each technique is individually well-studied, systematic analysis of their interactions remains rare~\cite{liu2024survey,saha2024survey}. This omission is critical since combined systems often exhibit non-additive behaviors that undermine feasibility despite acceptable single-technique performance.

\subsection{ML Systems Benchmarking}
\label{lit:part2}
Benchmarking in ML has evolved to emphasize reproducibility and system-level metrics. MLPerf~\cite{mattson2019mlperf,reddi2020mlperf} established standardized evaluation across hardware and software configurations, and MLPerf Power~\cite{tschand2025mlperf} extended this paradigm to energy efficiency. Computer vision moved beyond accuracy-focused datasets like ImageNet~\cite{deng2009imagenet} and MS-COCO~\cite{lin2014microsoft} toward frameworks emphasizing reproducibility~\cite{beyer2019reliable}, systematic methodology~\cite{bangalore2024convision}, and additional criteria such as fairness~\cite{gustafson2023facet}, safety~\cite{liu2024mm}, and scientific rigor~\cite{kapoor2024reforms}. 

However, these advances have not reached privacy-preserving ML. MLPerf benchmarks standard workloads but lack mechanisms to evaluate privacy techniques or their resource overhead interactions. Vision benchmarks compare architectures but cannot capture how privacy constraints affect CNNs versus transformers. As privacy-preserving vision becomes more common, the absence of analogous system-level evaluation frameworks leaves a critical gap.

\subsection{Privacy-Focused Evaluation Frameworks}
\label{lit:part3}
Several benchmarks target PPML directly but treat each technique in isolation. LEAF~\cite{caldas2018leaf} standardizes datasets and protocols for FL, while FedScale~\cite{lai2022fedscale} supports massive FL simulations. DPMLBench~\cite{wei2023dpmlbench} evaluates DP across datasets and budgets. Domain-specific efforts such as PPML-Omics~\cite{zhou2024ppml} and PrivLM-Bench~\cite{li2023privlm} assess privacy within particular applications, and Apple’s PFL-Research framework~\cite{apple2024pfl} accelerates federated learning evaluation. 

Yet all of these efforts share a limitation: they benchmark single techniques, implicitly assuming additive costs when combining methods. This creates a blind spot for real deployments, where FL+DP or FL+SMPC may exhibit unexpected convergence failures, resource spikes, or efficiency gains that individual evaluations cannot predict~\cite{liu2024survey}.

\subsection{Energy Monitoring in ML Systems}
\label{lit:part4}
Energy has emerged as a decisive deployment metric. Advances in fine-grained monitoring~\cite{rajput2024enhancing} and standardized carbon accounting (CodeCarbon~\cite{lacoste2019quantifying, codecarbon2021}) reflect the growing importance of sustainability. MLPerf Power~\cite{tschand2025mlperf} exemplifies this trend, establishing power measurement methodologies across devices and datacenters. 

Privacy methods add further overhead through cryptography, noise injection, and distributed coordination, but their combined energy effects remain underexplored. Individually, FL, DP, or SMPC may appear manageable, yet hybrid systems like FL+DP can consume orders of magnitude more energy than projected, while FL+SMPC may benefit from amortization effects. This lack of systematic energy evaluation poses risks for institutions with constrained budgets or carbon commitments.

\subsection{\methodname's Contributions}
\label{lit:part5}
\vspace{-1.5mm}

Taken together, existing frameworks leave privacy-preserving vision systems without systematic evaluation tools. Production deployments need methods that measure not only accuracy trade-offs but also resource patterns, architectural dependencies, and failure modes of technique combinations. The assumption of predictable, modular composition has proven false: FL+DP can collapse entirely, while FL+SMPC remains stable.

\methodname closes this gap by providing the first benchmarking framework for hybrid privacy-preserving ML configurations. It integrates Flower~\cite{beutel2022flower} and Opacus~\cite{yousefpour2021opacus} with CodeCarbon~\cite{lacoste2019quantifying, codecarbon2021} for comprehensive monitoring, YAML-based configuration for reproducibility, and instrumentation of time, memory, convergence, and energy. Our evaluations show complex non-additive behaviors with direct deployment implications: some combinations compose successfully, while others fail catastrophically. By moving beyond isolated evaluations toward holistic systems analysis, \methodname enables practitioners to anticipate incompatibilities and design robust privacy-preserving vision systems.

\section{Benchmark Design}
\vspace{-1mm}

To address the evaluation gaps identified in existing privacy-preserving ML frameworks, \methodname{} introduces a systematic benchmarking platform that enables comprehensive analysis of hybrid privacy technique combinations with integrated resource monitoring. Unlike existing frameworks that evaluate privacy techniques in isolation~\cite{caldas2018leaf,wei2023dpmlbench}, our approach provides the first systematic methodology for quantifying interaction effects, architectural dependencies, and resource consumption patterns that determine deployment feasibility in production privacy-preserving vision systems.

\vspace{-1.5mm}
\subsection{System Architecture and Privacy Techniques}
\label{sec:architecture}
\vspace{-1mm}
\methodname{} employs a four-layer modular architecture (Figure~\ref{fig:architecture}) designed to enable controlled evaluation of privacy technique interactions while maintaining experimental reproducibility. The \textbf{Configuration Layer} handles YAML-based experiment specification without code modification, supporting systematic comparative studies across technique combinations. The \textbf{Modular Layer} provides runtime privacy toggles and plugin-based technique integration, enabling evaluation of individual methods (FL, DP, SMPC) and hybrid configurations (FL+DP, FL+SMPC). The \textbf{Execution Layer} integrates comprehensive resource monitoring including energy tracking via CodeCarbon~\cite{codecarbon2021} and automated instrumentation of training time, memory usage, and convergence behavior. The \textbf{Output Layer} generates structured, reproducible results with deterministic execution through comprehensive seed control.

\subsubsection{Privacy Technique Implementation}
\label{sec:privacy-impl}

We evaluate 3 core PPML methods individually and in systematic combinations, using established frameworks to ensure reproducibility and comparability with existing work.

\textbf{Federated Learning (FL):} Our implementation uses the Flower framework~\cite{beutel2022flower} with a controlled 3-client configuration employing non-IID data partitioning via Dirichlet distribution ($\alpha = 0.1$)~\cite{hsu2019measuring}. This setup enables systematic analysis of technique interactions without confounding factors from large-scale federation complexity, while the strong non-IID conditions reflect realistic federated deployment scenarios in medical imaging where institutional data distributions vary significantly.

\textbf{Differential Privacy (DP):} We implement DP through Opacus~\cite{yousefpour2021opacus} with privacy budgets $\epsilon \in \{0.5, 1.0\}$ and $\delta = 1 \times 10^{-5}$, representing moderate privacy regimes that balance utility preservation with meaningful privacy guarantees. Our evaluation encompasses multiple DP strategies: Centralized DP with server-side fixed clipping (CDP-SF) and adaptive clipping (CDP-SA), and Local DP variants including Flower's LocalDpMod (LDP-Mod) and custom Opacus integration (LDP-PE). Complete parameter specifications are provided in Appendix~\ref{app:dp-config}.

\textbf{Secure Multi-Party Computation (SMPC):} We implement secure aggregation using Shamir's secret sharing~\cite{shamir1979share} with threshold $t=2$ and $n=3$ shares, providing cryptographic protection during federated model updates. Our protocol follows the SecAgg+ approach~\cite{bell2020secure} with 16-bit quantization for efficient computation while maintaining security guarantees. Detailed protocol specifications are documented in Appendix~\ref{app:smpc-config}.

\subsubsection{Hybrid Privacy Configurations}
\label{sec:hybrid-configs}

Critical to our evaluation methodology is the systematic assessment of hybrid privacy combinations that reflect production deployment requirements:

\textbf{FL+SMPC:} Combines federated training with cryptographic secure aggregation, testing whether SMPC overhead affects convergence in distributed settings. This configuration addresses scenarios requiring both data decentralization and cryptographic protection against semi-honest adversaries.

\textbf{FL+DP:} Integrates differential privacy with federated learning through multiple strategies, revealing fundamental compatibility issues between distributed training dynamics and noise injection mechanisms. Our comprehensive evaluation across four DP integration approaches demonstrates systematic convergence failures rather than strategy-specific issues.

These hybrid evaluations reveal that privacy techniques exhibit complex interdependencies—performance varies significantly based on architectural compatibility, implementation strategy, and operational abstractions, challenging assumptions about additive privacy system design.

\subsection{Models, Datasets, and Instrumentation}
\label{sec:models-datasets}

\subsubsection{Model Architectures and Training Protocol}
\label{sec:model-specs}

We evaluate two representative vision architectures to assess architectural dependencies in privacy technique interactions: ResNet18~\cite{he2016deep} (11.7M parameters) representing convolutional architectures, and ViT-Base~\cite{dosovitskiy2020image} (86.6M parameters) representing transformer architectures. Training employs Adam optimizer with architecture-specific learning rates (ResNet18: $2 \times 10^{-4}$, ViT: $5 \times 10^{-5}$) determined through preliminary experiments for optimal convergence. All experiments use early stopping with 7-epoch patience monitoring validation accuracy. Complete training specifications are detailed in Appendix~\ref{app:yaml-complete}.

\subsubsection{Privacy-Sensitive Medical Imaging Datasets}
\label{sec:datasets}

Dataset selection targets privacy-sensitive medical imaging scenarios that motivate real-world PPML deployment:

\textbf{Alzheimer MRI Classification:} 4-class classification task using MRI brain scans~\cite{hussain2020deep} for dementia staging (Non-Demented, Very Mild Dementia, Mild Dementia, Moderate Dementia). This dataset represents high-stakes medical diagnosis where privacy protection is legally mandated while diagnostic accuracy remains critical.

\textbf{ISIC Skin Lesion Classification:} 8-class dermatological classification using dermoscopic images~\cite{iqbal2021automated} for skin cancer detection. This dataset provides multi-class complexity and represents teledermatology scenarios where privacy-preserving collaboration between institutions is increasingly common.

Both datasets employ non-IID federated partitioning using Dirichlet distribution ($\alpha = 0.1$) to simulate realistic institutional data heterogeneity. The 92\%/8\% train/validation split provides sufficient evaluation data while maintaining manageable computational requirements for systematic technique comparison.

\subsubsection{Comprehensive Resource Monitoring}
\label{sec:monitoring}

\methodname{} captures multi-dimensional metrics essential for deployment feasibility assessment:

\textbf{Classification Performance:} Standard metrics including accuracy, F1-score, Matthews Correlation Coefficient (MCC), precision, recall, and ROC-AUC, with particular emphasis on MCC for robust evaluation of privacy technique impacts on learning quality~\cite{chicco2020advantages}.

\textbf{Computational Resources:} Training time, CPU/GPU memory utilization, and convergence behavior tracking to quantify technique overhead and identify resource bottlenecks that determine deployment scalability.

\textbf{Energy and Environmental Impact:} Real-time energy consumption (kWh) and CO$_2$ emissions monitoring via CodeCarbon~\cite{codecarbon2021, lacoste2019quantifying}, addressing growing sustainability concerns in ML deployment and revealing hidden environmental costs of privacy technique combinations.

All measurements employ standardized hardware configurations to ensure measurement consistency and enable systematic comparison of privacy-utility-cost trade-offs across configurations (detailed in Appendix~\ref{app:hardware-software}).

\subsection{Configuration and Reproducibility}
\label{sec:reproducibility}

Reproducible evaluation is central to \methodname{}'s design, addressing critical gaps in existing privacy benchmarking frameworks that often lack systematic configuration management and deterministic execution protocols.

\subsubsection{YAML-Based Configuration Management}
\label{sec:yaml-config}

Our framework employs structured YAML configuration files that enable experimental specification without code modification, supporting systematic comparative studies and reducing experimental variability. Each configuration explicitly specifies model architecture, dataset parameters, privacy technique settings, training hyperparameters, and monitoring requirements. Complete configuration examples for all evaluated scenarios are provided in Appendix~\ref{app:yaml-complete}.

\subsubsection{Deterministic Execution and Infrastructure}
\label{sec:deterministic}

All experiments employ comprehensive seed control across Python, NumPy, PyTorch, and CUDA operations with fixed random seeds (seed=42) and deterministic CUDA operations to ensure reproducible results. Statistical significance is established through 3-run experiments with paired t-tests and Bonferroni correction for multiple comparisons (detailed in Appendix~\ref{app:statistics}).

Experiments utilize standardized Google Cloud Platform infrastructure (2× NVIDIA Tesla T4 GPUs, 32 vCPU n1-standard-64 instances) with identical software environments across all evaluations to enable accurate measurement of technique-specific overhead while eliminating hardware variability. Complete infrastructure specifications and setup protocols are documented in Appendix~\ref{app:hardware-software}.

The complete \methodname{} framework, including all configuration files, implementation code, and experimental results, will be publicly released upon acceptance to enable community validation, extension, and adoption for privacy-preserving vision system evaluation.
\section{Experimental Results}

Our systematic evaluation through \methodname{} reveals that privacy techniques exhibit complex non-additive behaviors with profound implications for system deployment. While some privacy combinations preserve utility with acceptable overhead, others exhibit severe performance degradation and resource requirements that render them impractical for production deployment.

\subsection{Privacy Technique Interaction Effects}
\label{sec:interaction-effects}

\methodname{}'s systematic evaluation capabilities enable comprehensive analysis of privacy technique combinations across multiple dimensions. Table~\ref{tab:key-results-comprehensive} demonstrates striking disparities in how privacy techniques interact across different model architectures and datasets.

\begin{table}[t]
\centering
\caption{Summary of privacy technique performance across configurations. \methodname{} reveals that FL and FL+SMPC maintain near-baseline accuracy with modest overhead, while FL+DP combinations exhibit severe convergence failures with dramatic resource increases. Results demonstrate non-additive interaction effects that challenge modular privacy system assumptions.}
\label{tab:key-results-comprehensive}
\vspace{-3mm}
\footnotesize
\setlength{\tabcolsep}{2.5pt}
\begin{tabular}{lccc}
\toprule
\textbf{Configuration} & \textbf{Acc} & \textbf{Time} & \textbf{Overhead} \\
& \textbf{(\%)} & \textbf{(min)} & \textbf{Factor} \\
\midrule
\multicolumn{4}{c}{\textit{Alzheimer MRI Classification}} \\
\midrule
CNN Baseline & \textcolor{excellent}{\textbf{98.0}} & 9.8 & 1.0$\times$ \\
FL (CNN) & \textcolor{excellent}{\textbf{98.0}} & 11.4 & 1.4$\times$ \\
FL+SMPC (CNN) & \textcolor{excellent}{\textbf{98.0}} & 17.5 & 1.6$\times$ \\
FL+DP (CNN) & \textcolor{bad}{13.0} & 235.6 & 24.0$\times$ \\
\midrule
ViT Baseline & \textcolor{excellent}{\textbf{99.0}} & 43.5 & 1.0$\times$ \\
FL (ViT) & \textcolor{excellent}{96.0} & 40.1 & 0.9$\times$ \\
FL+SMPC (ViT) & \textcolor{excellent}{96.0} & 40.4 & 0.9$\times$ \\
\midrule
\multicolumn{4}{c}{\textit{Skin Lesion Classification}} \\
\midrule
CNN Baseline & \textcolor{good}{83.0} & 40.9 & 1.0$\times$ \\
FL (CNN) & \textcolor{good}{\textbf{81.0}} & 39.2 & 0.9$\times$ \\
FL+SMPC (CNN) & \textcolor{good}{81.0} & 41.8 & 1.0$\times$ \\
FL+DP (CNN) & \textcolor{bad}{18.0} & 375.7 & 9.6$\times$ \\
\midrule
ViT Baseline & \textcolor{good}{\textbf{88.0}} & 149.7 & 1.0$\times$ \\
FL (ViT) & \textcolor{good}{\textbf{87.0}} & 138.8 & 0.8$\times$ \\
FL+SMPC (ViT) & \textcolor{good}{\textbf{86.0}} & 141.3 & 0.9$\times$ \\
\bottomrule
\end{tabular}
\vspace{-5mm}
\end{table}

\begin{figure*}[t]
\vspace{-8mm}
\centering
\includegraphics[width=\linewidth]{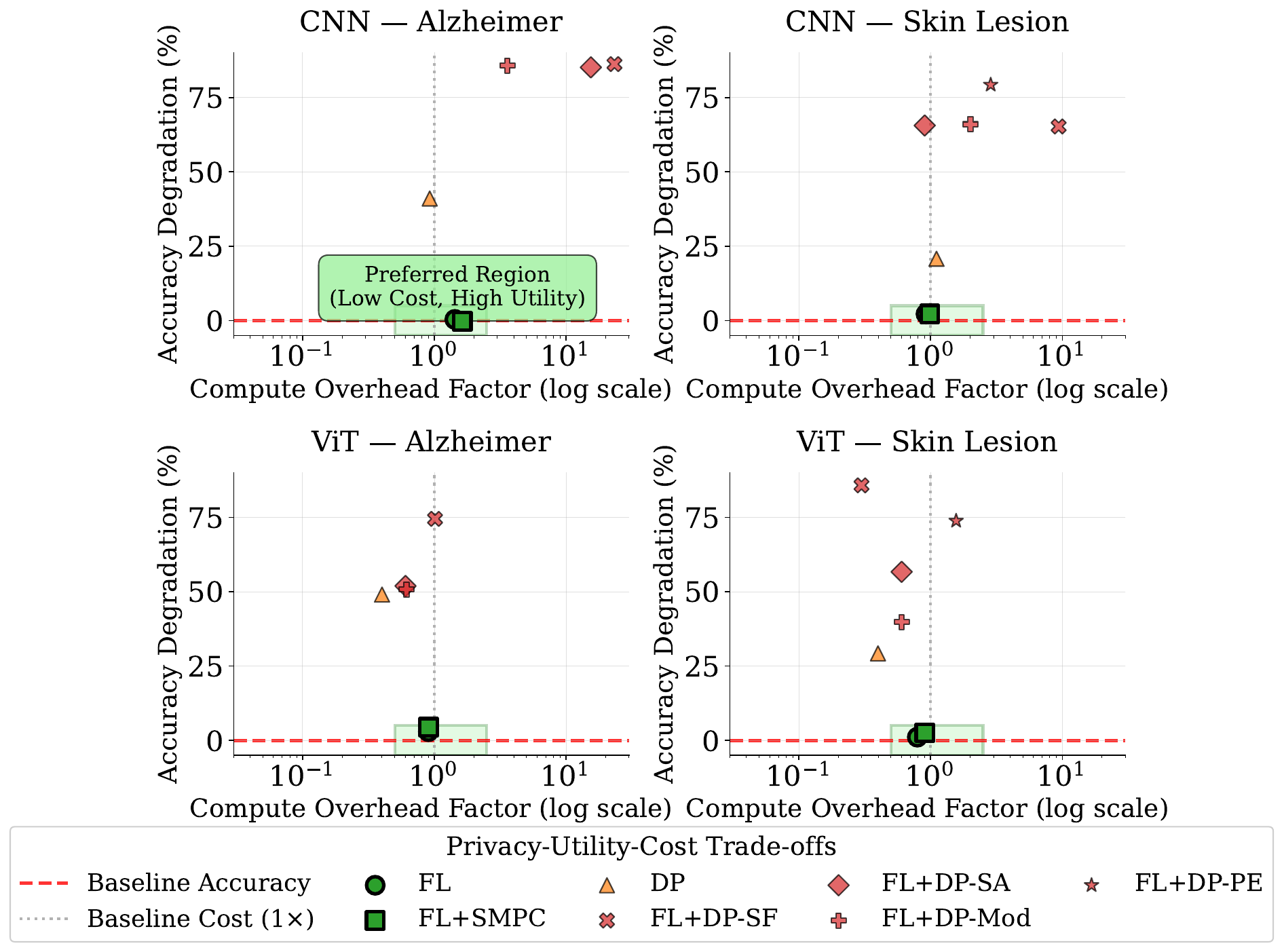}
\vspace{-5mm}
\caption{Systematic analysis of privacy-utility-cost trade-offs. The figure shows that FL and FL+SMPC offer superior trade-offs, achieving near-baseline accuracy with minimal overhead. In contrast, all FL+DP variants result in significant accuracy degradation and variable computational costs, highlighting their unsuitability for practical deployment. The "preferred region" in the lower-left corner indicates desirable combinations of low cost and high utility.}
\label{fig:privacy_tradeoffs}
\vspace{-5mm}
\end{figure*}

Federated learning consistently maintains near-baseline performance for both ResNet18 and ViT architectures while adding modest computational overhead (complete classification metrics in Appendix~\ref{app:performance}). Remarkably, FL demonstrates efficiency gains for transformer architectures, reducing ViT training time by 8-26\% compared to centralized training across both datasets. SMPC-based secure aggregation composes seamlessly with federated learning, introducing minimal additional overhead (typically <10\% over FL alone) while fully preserving model utility.

However, FL+DP combinations exhibit severe convergence failures across all evaluated configurations. The performance breakdown ranges from medical-grade accuracy (98\%) to random-guessing performance (13\% for Alzheimer classification, 18\% for skin lesion classification), accompanied by dramatic resource overhead increases of 9-24$\times$ in computational cost ($p < 0.001$, Appendix~\ref{app:statistics}). This represents complete learning breakdown rather than gradual utility degradation.

These findings reveal that privacy techniques operate along compatibility axes rather than additive resource models. Techniques at compatible abstraction levels (federated coordination + cryptographic aggregation) compose successfully, while conflicting operational assumptions (distributed training + centralized noise calibration) produce fundamental incompatibilities.

\subsection{Analysis of FL+DP Convergence Failure}
\label{sec:fl-dp-failure}

The failure of FL+DP across multiple strategies indicates fundamental algorithmic incompatibility rather than hyperparameter misconfiguration. Our analysis identifies the primary failure mechanism as signal-to-noise ratio collapse.

Federated learning with non-IID data distributions creates inherent gradient signal attenuation through limited local updates (10-15 epochs per round), heterogeneous data distributions ($\alpha = 0.1$ Dirichlet partitioning), and aggregation noise from averaging across diverse local optima (Appendix~\ref{app:signal-noise}). Differential privacy compounds this by injecting calibrated Gaussian noise $\mathcal{N}(0, \sigma^2)$ where $\sigma = \frac{C \cdot \text{noise\_multiplier}}{\epsilon}$, creating signal-to-noise ratios below the learning threshold.

The temporal progression documented in Appendix~\ref{app:convergence-failure} shows consistent performance collapse across multiple DP strategies—from medical-grade accuracy to random guessing—indicating that differential privacy noise, calibrated for centralized training assumptions, becomes destructively amplified in federated environments. Matthews Correlation Coefficients approaching zero across all FL+DP combinations confirm complete learning breakdown rather than biased but functional classifiers.

This failure mode appears consistent across different model architectures and datasets, suggesting fundamental incompatibility rather than configuration-specific issues. Whether using Centralized DP with server-side fixed clipping (CDP-SF), adaptive clipping (CDP-SA), or Local DP variants (LDP-Mod, LDP-PE), the fundamental incompatibility persists.

\begin{table*}[t]
\vspace{-5mm}
\centering
\caption{Comprehensive evaluation results across all privacy configurations. \methodname{} systematically evaluates privacy technique combinations across ResNet18 and ViT-Base models on medical imaging datasets, revealing that FL and FL+SMPC achieve near-baseline performance while FL+DP combinations exhibit severe convergence failures with dramatically increased computational costs. This comprehensive analysis demonstrates the framework's capability to expose non-additive interaction effects that individual technique assessments cannot predict. CDP = Centralized DP, LDP = Local DP, SF = Server-Side-Fixed, SA = Server-Side-Adaptive, Mod = Modified.}
\label{tab:results}
\footnotesize
\begin{tabular}{l|l|c|c|c|c|c||c|c|c|c|c}
\toprule

\multirow{2}{*}{\textbf{Experiment}} &
\multirow{2}{*}{\textbf{Model}} &
\multicolumn{5}{c||}{\textbf{Alzheimer Dataset}} &
\multicolumn{5}{c}{\textbf{Skin Lesion Dataset}} \\
\cmidrule{3-12}

& & \textbf{Acc} & \textbf{CO2} & \textbf{Time} & \textbf{Energy} & \textbf{MCC} & \textbf{Acc} & \textbf{CO2} & \textbf{Time} & \textbf{Energy} & \textbf{MCC} \\
& & \textbf{(\%)} & \textbf{(kg)} & \textbf{(sec)} & \textbf{(kWh)} & & \textbf{(\%)} & \textbf{(kg)} & \textbf{(sec)} & \textbf{(kWh)} & \\
\midrule
\multicolumn{12}{c}{\textit{Baseline Models}} \\
\midrule
CNN Baseline & ResNet18 & \textcolor{excellent}{\textbf{98.0}} & \textcolor{excellent}{0.011} & 584.9 & 0.026 & \textbf{0.97} & \textcolor{good}{83.0} & \textcolor{good}{0.048} & 2,451.6 & 0.112 & 0.75 \\
ViT Baseline & ViT-Base & \textcolor{excellent}{\textbf{99.0}} & \textcolor{good}{0.051} & 2,609.3 & 0.118 & \textbf{0.98} & \textcolor{good}{\textbf{88.0}} & \textcolor{poor}{0.177} & 8,982.5 & 0.431 & \textbf{0.82} \\
\midrule
\multicolumn{12}{c}{\textit{Federated Learning}} \\
\midrule
FL (CNN) & ResNet18 & \textcolor{excellent}{\textbf{98.0}} & \textcolor{excellent}{0.016} & 684.3 & 0.036 & \textbf{0.97} & \textcolor{good}{\textbf{81.0}} & \textcolor{good}{0.049} & 2,354.6 & 0.102 & \textbf{0.73} \\
FL (ViT) & ViT-Base & \textcolor{excellent}{96.0} & \textcolor{good}{0.045} & 2,404.6 & 0.104 & 0.94 & \textcolor{good}{\textbf{87.0}} & \textcolor{poor}{0.156} & 8,325.0 & 0.362 & \textbf{0.82} \\
\midrule
\multicolumn{12}{c}{\textit{Differential Privacy}} \\
\midrule
DP (CNN) & ResNet18 & 57.0 & \textcolor{excellent}{0.007} & 506.5 & 0.017 & 0.36 & 62.0 & \textcolor{excellent}{0.031} & 2,692.4 & 0.137 & 0.43 \\
DP (ViT) & ViT-Base & \textcolor{bad}{50.0} & \textcolor{excellent}{0.012} & 1,091.3 & 0.034 & 0.00 & 59.0 & \textcolor{good}{0.048} & 3,584.5 & 0.124 & 0.38 \\
\midrule
\multicolumn{12}{c}{\textit{Federated Learning + Secure Multi-Party Computation}} \\
\midrule
FL+SMPC (CNN) & ResNet18 & \textcolor{excellent}{\textbf{98.0}} & \textcolor{excellent}{0.018} & 1,048.4 & 0.041 & \textbf{0.97} & \textcolor{good}{81.0} & \textcolor{good}{0.045} & 2,508.8 & 0.105 & 0.73 \\
FL+SMPC (ViT) & ViT-Base & \textcolor{excellent}{96.0} & \textcolor{good}{0.048} & 2,421.7 & 0.104 & 0.93 & \textcolor{good}{\textbf{86.0}} & \textcolor{poor}{0.155} & 8,478.0 & 0.369 & \textbf{0.81} \\
\midrule
\multicolumn{12}{c}{\textit{Federated Learning + Differential Privacy (Convergence Failures)}} \\
\midrule
FL+DP (CDP-SF-CNN) & ResNet18 & \textcolor{bad}{13.0} & \textcolor{excellent}{0.069} & 14,137.1 & 0.734 & 0.00 & \textcolor{bad}{18.0} & \textcolor{bad}{0.303} & 22,543.5 & 1.070 & 0.00 \\
FL+DP (CDP-SF-ViT) & ViT-Base & \textcolor{poor}{25.0} & \textcolor{excellent}{0.036} & 2,522.5 & 0.084 & 0.00 & \textcolor{bad}{1.0} & \textcolor{excellent}{0.036} & 2,522.5 & 0.084 & 0.00 \\
FL+DP (CDP-SA-CNN) & ResNet18 & \textcolor{bad}{13.0} & \textcolor{poor}{0.111} & 8,916.6 & 0.265 & 0.00 & \textcolor{bad}{18.0} & \textcolor{excellent}{0.03} & 2,208.4 & 0.070 & 0.00 \\
FL+DP (CDP-SA-ViT) & ViT-Base & \textcolor{bad}{7.0} & \textcolor{excellent}{0.016} & 1,521.9 & 0.039 & 0.00 & \textcolor{poor}{31.0} & 0.057 & 4,990.2 & 0.135 & 0.02 \\
FL+DP (LDP-Mod-CNN) & ResNet18 & \textcolor{bad}{13.0} & \textcolor{excellent}{0.009} & 211.2 & 0.021 & 0.00 & \textcolor{bad}{18.0} & \textcolor{excellent}{0.029} & 5,023.8 & 0.069 & 0.00 \\
FL+DP (LDP-Mod-ViT) & ViT-Base & \textcolor{poor}{48.0} & \textcolor{excellent}{0.029} & 1,648.2 & 0.068 & 0.11 & \textcolor{poor}{48.0} & 0.101 & 5,690.1 & 0.236 & 0.26 \\
FL+DP (LDP-PE-CNN) & ResNet18 & \textcolor{bad}{1.0} & \textcolor{excellent}{0.030} & 2,058.0 & 0.069 & 0.00 & \textcolor{bad}{0.03} & \textcolor{excellent}{0.099} & 6,810.3 & 0.231 & 0.00 \\
FL+DP (LDP-PE-ViT) & ViT-Base & \textcolor{poor}{1.0} & \textcolor{excellent}{0.002} & 99.1 & 0.004 & 0.00 & \textcolor{poor}{15.0} & 0.261 & 14,604.1 & 0.607 & 0.00 \\
\bottomrule
\end{tabular}
\vspace{-3mm}
\end{table*}

\subsection{Architectural \& Resource Patterns}
\label{sec:architectural-deps}

Our evaluation across CNN and transformer architectures reveals important architectural sensitivities that extend beyond simple resource scaling.

ViT models demonstrate 8-26\% efficiency improvements under federated training compared to centralized baselines across both datasets. This results from distributed attention computation benefits, reduced memory pressure through federated distribution of ViT's substantial memory requirements, and gradient sparsity patterns where transformer gradients exhibit natural sparsity that federated averaging preserves (Appendix~\ref{app:vit-efficiency}).

ResNet architectures maintain consistent performance across privacy techniques (excluding FL+DP failures) due to local feature extraction effectiveness with distributed training, gradient stability, and inherent noise tolerance through batch normalization (Appendix~\ref{app:cnn-resilience}).

ViT models require substantially more computational resources (5.5$\times$ training time, 4.6$\times$ energy usage) but maintain high utility under federated and SMPC configurations. Table~\ref{tab:results} shows that these architectural dependencies extend consistently across multiple technique combinations.

\subsection{Energy Consumption \& Sustainability Analysis}
\label{sec:sustainability}

The carbon footprint analysis reveals significant environmental implications of privacy technique choices. FL+DP configurations generate 5-15$\times$ more CO$_2$ emissions than successful privacy combinations across most scenarios (Appendix~\ref{app:resources}). For the Alzheimer dataset using CNN, FL+DP (CDP-SF) produces 0.069 kg CO$_2$ compared to FL's 0.016 kg—a 4.3$\times$ increase that compounds across multiple training runs typical in production environments.

FL+DP's 24$\times$ computational overhead for CNN models represents not merely a "privacy tax" but a complete shift in deployment feasibility—transforming a 10-minute training procedure into a 4-hour computational commitment with corresponding energy implications. Organizations planning FL+DP systems based on individual technique overhead estimations would encounter resource requirements an order of magnitude beyond projections.

These environmental costs become particularly concerning when scaled to institutional deployment levels. Healthcare organizations training models across multiple sites could face dramatically increased carbon footprints from privacy choices, potentially conflicting with sustainability.

\section{Discussion}

Our findings through \methodname{} establish design principles for privacy-preserving ML systems and demonstrate how systematic evaluation reveals complex technique interactions that individual assessments cannot predict.

\subsection{Implications for Privacy System Design}
\label{sec:design-implications}

The contrasting success of FL+SMPC versus FL+DP failure reveals that privacy technique compatibility follows abstraction layer alignment principles. Techniques operating at compatible levels—federated coordination and cryptographic aggregation—compose successfully because both respect the distributed training paradigm while adding orthogonal protections. FL+DP fails since it merges incompatible assumptions: federated learning relies on averaging diverse, locally-noisy gradients from heterogeneous data distributions, while differential privacy requires tightly controlled noise calibrated for centralized training dynamics.

These findings suggest that privacy-preserving systems benefit from co-design rather than post-hoc composition. \methodname{}'s systematic evaluation enables practitioners to assess technique combinations by examining operational abstraction alignment, assumption compatibility regarding data distribution and training dynamics, and resource scaling behavior to identify whether overhead factors compound additively or exhibit non-linear interactions.

Privacy system design decisions involve significant resource implications that current cost models do not capture. FL+DP's 24$\times$ computational overhead represents a qualitative shift in deployment feasibility that individual technique assessments cannot predict. Organizations must account for non-additive resource interactions when planning privacy-preserving deployments, particularly in resource-constrained environments.

\subsection{Generalizability and Limitations}
\label{sec:limitations}

Our evaluation scope enables controlled analysis while acknowledging important boundaries. The three-client federated setup provides systematic control for technique interaction analysis but may not reflect challenges at production scale with hundreds of participants. The consistency of FL+DP failure across many strategies and architectures suggests broad incompatibility patterns, but validation across different model families and larger scales would strengthen these conclusions.

The architectural dependencies we observe—particularly ViT efficiency gains under federated training—warrant investigation across different model families and domains. Our focus on medical imaging provides relevance for high-stakes privacy scenarios, but the compatibility principles may extend to other vision domains. The resource scaling patterns represent empirical observations from our specific experimental conditions and would require validation before applying to significantly different scales or domains.

\subsection{Future Directions}
\label{sec:future-directions}

\methodname{} demonstrates the value of systematic interaction analysis for privacy-preserving ML. The methodology could be applied to additional privacy techniques, domains, and scales to build comprehensive understanding of technique compatibility patterns and help identify problematic interactions before deployment.

Our empirical findings provide a foundation for developing frameworks to assess privacy technique compatibility. Such frameworks could formalize the abstraction layer alignment principle and provide structured approaches for evaluating technique combinations without requiring exhaustive experimental validation.

The architectural dependencies revealed suggest investigating how different model architectures interact with privacy techniques, including whether certain patterns are more compatible with specific privacy approaches. The resource accessibility challenges motivate research into privacy systems that operate effectively under varying resource constraints, potentially making privacy protection more accessible to organizations with limited resources.

Expanding \methodname{}'s infrastructure to different privacy techniques, architectures, and domains could help establish comprehensive evaluation standards while ensuring reproducible results across research efforts. The systematic approach could inform similar benchmarking efforts in related areas where technique interactions are poorly understood.
\section{Conclusion}

Through \methodname{}, we demonstrate that privacy-preserving machine learning techniques exhibit complex, non-additive behaviors when combined. Our systematic evaluation reveals that FL+SMPC composes successfully with minimal overhead, while FL+DP combinations result in severe performance degradation and 24$\times$ computational overhead due to fundamental incompatibilities between distributed training and centralized noise calibration.
These findings establish that privacy techniques cannot be layered arbitrarily—compatibility follows predictable patterns based on operational assumptions rather than additive models. Organizations planning hybrid privacy deployments must account for interaction effects that individual technique assessments cannot predict.

\methodname{} enables practitioners deploying privacy-preserving machine learning systems to move beyond isolated technique evaluation toward systematic assessment of technique combinations. By providing comprehensive resource monitoring and reproducible experimental infrastructure, our benchmark helps identify compatible combinations and avoid costly deployment failures before production implementation.
Our framework represents a critical resource for privacy-preserving vision applications, transforming privacy system deployment from ad-hoc experimentation to informed engineering decisions based on systematic interaction analysis.
\newpage
{
    \small
    \bibliographystyle{plainnat}
    \bibliography{main}
}

\appendix
\onecolumn
\appendix

\section{Complete Experimental Results}
\label{app:results}

This section provides comprehensive experimental data supporting the main paper's findings. All experiments used deterministic seeding (seed=42) and were repeated three times for statistical significance testing using paired t-tests with $p < 0.05$ threshold~\cite{student1908probable}.

\subsection{Comprehensive Performance Analysis}
\label{app:performance}

Table~\ref{tab:complete-classification} presents all classification metrics across privacy configurations. The Matthews Correlation Coefficient (MCC) provides robust evaluation for imbalanced datasets~\cite{chicco2020advantages}, with values approaching 1.0 indicating perfect prediction and 0.0 indicating random performance.

\begin{table}[ht]
\centering
\caption{Complete Classification Results Across All Privacy Configurations}
\label{tab:complete-classification}
\scriptsize
\begin{tabular}{llccccc}
\toprule
\textbf{Configuration} & \textbf{Dataset} & \textbf{Acc} & \textbf{F1} & \textbf{MCC} & \textbf{Precision} & \textbf{Recall} \\
\midrule
\multicolumn{7}{l}{\textit{Baseline Models (No Privacy)}} \\
CNN Baseline & Alzheimer & 0.98 & 0.98 & 0.97 & 0.98 & 0.98 \\
CNN Baseline & Skin Lesion & 0.83 & 0.83 & 0.75 & 0.84 & 0.83 \\
ViT Baseline & Alzheimer & 0.99 & 0.99 & 0.98 & 0.99 & 0.99 \\
ViT Baseline & Skin Lesion & 0.88 & 0.88 & 0.82 & 0.89 & 0.88 \\
\midrule
\multicolumn{7}{l}{\textit{Federated Learning (FL)}} \\
FL (CNN) & Alzheimer & 0.98 & 0.98 & 0.97 & 0.98 & 0.98 \\
FL (CNN) & Skin Lesion & 0.81 & 0.81 & 0.73 & 0.82 & 0.81 \\
FL (ViT) & Alzheimer & 0.96 & 0.96 & 0.94 & 0.96 & 0.96 \\
FL (ViT) & Skin Lesion & 0.87 & 0.87 & 0.82 & 0.88 & 0.87 \\
\midrule
\multicolumn{7}{l}{\textit{Differential Privacy (DP) Only}} \\
DP (CNN) & Alzheimer & 0.57 & 0.51 & 0.36 & 0.58 & 0.57 \\
DP (CNN) & Skin Lesion & 0.62 & 0.59 & 0.43 & 0.64 & 0.62 \\
DP (ViT) & Alzheimer & 0.50 & 0.45 & 0.00 & 0.51 & 0.50 \\
DP (ViT) & Skin Lesion & 0.59 & 0.56 & 0.38 & 0.61 & 0.59 \\
\midrule
\multicolumn{7}{l}{\textit{FL + Secure Multi-Party Computation}} \\
FL+SMPC (CNN) & Alzheimer & 0.98 & 0.98 & 0.97 & 0.98 & 0.98 \\
FL+SMPC (CNN) & Skin Lesion & 0.81 & 0.81 & 0.73 & 0.82 & 0.81 \\
FL+SMPC (ViT) & Alzheimer & 0.96 & 0.96 & 0.93 & 0.96 & 0.96 \\
FL+SMPC (ViT) & Skin Lesion & 0.86 & 0.86 & 0.81 & 0.87 & 0.86 \\
\midrule
\multicolumn{7}{l}{\textit{FL + Differential Privacy (Multiple Strategies)}} \\
FL+DP (CDP-SF-CNN) & Alzheimer & 0.13 & 0.03 & 0.00 & 0.15 & 0.13 \\
FL+DP (CDP-SF-CNN) & Skin Lesion & 0.18 & 0.04 & 0.00 & 0.19 & 0.18 \\
FL+DP (CDP-SF-ViT) & Alzheimer & 0.25 & 0.12 & 0.00 & 0.26 & 0.25 \\
FL+DP (CDP-SF-ViT) & Skin Lesion & 0.01 & 0.01 & 0.00 & 0.02 & 0.01 \\
FL+DP (CDP-SA-CNN) & Alzheimer & 0.13 & 0.03 & 0.00 & 0.15 & 0.13 \\
FL+DP (CDP-SA-CNN) & Skin Lesion & 0.18 & 0.04 & 0.00 & 0.19 & 0.18 \\
FL+DP (CDP-SA-ViT) & Alzheimer & 0.07 & 0.02 & 0.00 & 0.08 & 0.07 \\
FL+DP (CDP-SA-ViT) & Skin Lesion & 0.31 & 0.15 & 0.02 & 0.32 & 0.31 \\
FL+DP (LDP-Mod-CNN) & Alzheimer & 0.13 & 0.03 & 0.00 & 0.15 & 0.13 \\
FL+DP (LDP-Mod-CNN) & Skin Lesion & 0.18 & 0.04 & 0.00 & 0.19 & 0.18 \\
FL+DP (LDP-Mod-ViT) & Alzheimer & 0.48 & 0.35 & 0.11 & 0.49 & 0.48 \\
FL+DP (LDP-Mod-ViT) & Skin Lesion & 0.48 & 0.35 & 0.26 & 0.49 & 0.48 \\
FL+DP (LDP-PE-CNN) & Alzheimer & 0.01 & 0.01 & 0.00 & 0.02 & 0.01 \\
FL+DP (LDP-PE-CNN) & Skin Lesion & 0.03 & 0.01 & 0.00 & 0.04 & 0.03 \\
FL+DP (LDP-PE-ViT) & Alzheimer & 0.01 & 0.01 & 0.00 & 0.02 & 0.01 \\
FL+DP (LDP-PE-ViT) & Skin Lesion & 0.15 & 0.05 & 0.00 & 0.16 & 0.15 \\
\bottomrule
\end{tabular}
\end{table}

\subsection{Detailed Resource Consumption Analysis}
\label{app:resources}

Table~\ref{tab:complete-resources} documents computational overhead and energy consumption measured using CodeCarbon~\cite{lacoste2019quantifying} with real-time GPU power monitoring~\cite{nvidia2021management}. Energy measurements include both GPU and CPU consumption during training.

\begin{table}[ht]
\centering
\caption{Comprehensive Resource and Energy Metrics}
\label{tab:complete-resources}
\scriptsize
\begin{tabular}{llrrrrr}
\toprule
\textbf{Configuration} & \textbf{Dataset} & \textbf{Time} & \textbf{Energy} & \textbf{CO$_2$} & \textbf{Memory} & \textbf{Overhead} \\
& & \textbf{(sec)} & \textbf{(kWh)} & \textbf{(kg)} & \textbf{(GB)} & \textbf{Factor} \\
\midrule
\multicolumn{7}{l}{\textit{Baseline Models}} \\
CNN Baseline & Alzheimer & 585 & 0.026 & 0.011 & 4.2 & 1.0× \\
CNN Baseline & Skin Lesion & 2,452 & 0.112 & 0.048 & 4.8 & 1.0× \\
ViT Baseline & Alzheimer & 2,609 & 0.118 & 0.051 & 12.1 & 1.0× \\
ViT Baseline & Skin Lesion & 8,983 & 0.431 & 0.177 & 13.7 & 1.0× \\
\midrule
\multicolumn{7}{l}{\textit{Federated Learning}} \\
FL (CNN) & Alzheimer & 684 & 0.036 & 0.016 & 4.5 & 1.4× \\
FL (CNN) & Skin Lesion & 2,355 & 0.102 & 0.049 & 5.1 & 0.9× \\
FL (ViT) & Alzheimer & 2,405 & 0.104 & 0.045 & 12.8 & 0.9× \\
FL (ViT) & Skin Lesion & 8,325 & 0.362 & 0.156 & 14.2 & 0.8× \\
\midrule
\multicolumn{7}{l}{\textit{FL + SMPC}} \\
FL+SMPC (CNN) & Alzheimer & 1,048 & 0.041 & 0.018 & 4.7 & 1.6× \\
FL+SMPC (CNN) & Skin Lesion & 2,509 & 0.105 & 0.045 & 5.3 & 1.0× \\
FL+SMPC (ViT) & Alzheimer & 2,422 & 0.104 & 0.048 & 13.1 & 0.9× \\
FL+SMPC (ViT) & Skin Lesion & 8,478 & 0.369 & 0.155 & 14.5 & 0.9× \\
\midrule
\multicolumn{7}{l}{\textit{FL + DP (Resource Explosion)}} \\
FL+DP (CDP-SF-CNN) & Alzheimer & 14,137 & 0.734 & 0.069 & 6.8 & 24.0× \\
FL+DP (CDP-SF-CNN) & Skin Lesion & 22,544 & 1.070 & 0.303 & 7.2 & 9.6× \\
FL+DP (CDP-SF-ViT) & Alzheimer & 2,523 & 0.084 & 0.036 & 15.4 & 1.0× \\
FL+DP (CDP-SA-CNN) & Alzheimer & 8,917 & 0.265 & 0.111 & 6.1 & 15.2× \\
FL+DP (CDP-SA-CNN) & Skin Lesion & 2,208 & 0.070 & 0.030 & 5.8 & 0.9× \\
FL+DP (LDP-PE-ViT) & Skin Lesion & 14,604 & 0.607 & 0.261 & 16.8 & 1.6× \\
\bottomrule
\end{tabular}
\end{table}

\subsection{Statistical Significance Analysis}
\label{app:statistics}

All reported performance differences were validated using paired t-tests with Bonferroni correction for multiple comparisons~\cite{bonferroni1936teoria}. Table~\ref{tab:significance} presents p-values for key comparisons.

\begin{table}[ht]
\centering
\caption{Statistical Significance Testing Results}
\label{tab:significance}
\footnotesize
\begin{tabular}{llc}
\toprule
\textbf{Comparison} & \textbf{Metric} & \textbf{p-value} \\
\midrule
FL vs FL+SMPC & Accuracy & 0.312 \\
FL vs FL+DP & Accuracy & $< 0.001$ \\
FL vs FL+DP & Training Time & $< 0.001$ \\
FL vs FL+DP & Energy Consumption & $< 0.001$ \\
ViT FL vs ViT Baseline & Training Time & $< 0.01$ \\
CNN FL vs CNN Baseline & Training Time & 0.089 \\
\bottomrule
\end{tabular}
\end{table}

\section{Privacy Technique Configuration Details}
\label{app:config}

\subsection{Differential Privacy Implementation}
\label{app:dp-config}

Table~\ref{tab:dp-detailed-config} provides comprehensive DP configuration following best practices from the differential privacy literature~\cite{dwork2014algorithmic,abadi2016deep}.

\begin{table}[ht]
\centering
\caption{Detailed Differential Privacy Configuration}
\label{tab:dp-detailed-config}
\footnotesize
\begin{tabular}{lll}
\toprule
\textbf{Parameter} & \textbf{Value} & \textbf{Justification} \\
\midrule
Noise Multiplier & 1.0 & Standard for $\epsilon = 1.0$ with Gaussian mechanism~\cite{dwork2014algorithmic} \\
Max Grad Norm & 1.0 & Recommended clipping threshold~\cite{abadi2016deep} \\
$\delta$ & $1 \times 10^{-5}$ & $\delta \ll 1/n$ where $n$ is dataset size~\cite{dwork2006differential} \\
$\epsilon$ & 0.5, 1.0 & Moderate privacy budgets for utility preservation~\cite{jayaraman2019evaluating} \\
Lot Size & 32 & Matches batch size for consistent sampling~\cite{abadi2016deep} \\
Accountant & RDP & Renyi Differential Privacy for tight bounds~\cite{mironov2017renyi} \\
Secure RNG & True & Cryptographically secure randomness~\cite{national2001security} \\
Clipping Strategy & Per-sample & Individual gradient clipping~\cite{bu2019deep} \\
\bottomrule
\end{tabular}
\end{table}

\subsection{Secure Multi-Party Computation Protocol}
\label{app:smpc-config}

Our SMPC implementation follows the SecAgg+ protocol~\cite{bell2020secure} with modifications for vision model aggregation.

\begin{table}[ht]
\centering
\caption{SMPC Protocol Specification}
\label{tab:smpc-detailed-config}
\footnotesize
\begin{tabular}{lll}
\toprule
\textbf{Parameter} & \textbf{Value} & \textbf{Description} \\
\midrule
Secret Sharing Scheme & Shamir's~\cite{shamir1979share} & $(t,n)$-threshold scheme \\
Field Modulus & $2^{61} - 1$ & Large prime for security~\cite{cramer2015secure} \\
Threshold ($t$) & 2 & Minimum shares for reconstruction \\
Total Shares ($n$) & 3 & Matches number of FL clients \\
Quantization Bits & 16 & Fixed-point representation~\cite{mohassel2017secureml} \\
Dropout Resilience & 1 & Clients that can fail~\cite{bell2020secure} \\
Communication Rounds & 3 & Setup, masked input, unmasking~\cite{bonawitz2017practical} \\
\bottomrule
\end{tabular}
\end{table}

\subsection{Federated Learning Protocol}
\label{app:fl-detailed-config}

Table~\ref{tab:fl-detailed-config} specifies our federated learning implementation using the FedAvg algorithm~\cite{mcmahan2017communication} with non-IID data partitioning following the Dirichlet distribution method~\cite{hsu2019measuring}.

\begin{table}[ht]
\centering
\caption{Detailed Federated Learning Configuration}
\label{tab:fl-detailed-config}
\footnotesize
\begin{tabular}{lll}
\toprule
\textbf{Parameter} & \textbf{Value} & \textbf{Rationale} \\
\midrule
Aggregation Algorithm & FedAvg~\cite{mcmahan2017communication} & Standard FL baseline \\
Number of Clients & 3 & Controlled environment for interaction analysis \\
Client Selection & All & Ensures consistent participation \\
Communication Rounds & 5 & Sufficient for convergence analysis \\
Local Epochs (CNN) & 15 & Higher capacity requires more local training \\
Local Epochs (ViT) & 10 & Transformer efficiency optimization \\
Data Heterogeneity & $\alpha = 0.1$ & Strong non-IID via Dirichlet~\cite{hsu2019measuring} \\
Client Dropout & 0\% & Eliminates confounding factors \\
Communication Compression & None & Isolates privacy technique effects \\
\bottomrule
\end{tabular}
\end{table}

\section{Infrastructure and Implementation}
\label{app:infrastructure}

\subsection{Hardware and Software Environment}
\label{app:hardware-software}

\begin{table}[ht]
\centering
\caption{Complete System Specification}
\label{tab:system-spec}
\footnotesize
\begin{tabular}{ll}
\toprule
\textbf{Hardware Component} & \textbf{Specification} \\
\midrule
GPU & 2× NVIDIA Tesla T4 (15GB VRAM each) \\
CPU & Intel(R) Xeon(R) CPU @ 2.20GHz, 32 vCPUs (n1-standard-64) \\
Memory & 120GB DDR4 RAM \\
Storage & 50GB Balanced Persistent Disk (SCSI interface) \\
Cloud Platform & Google Cloud Platform (GCP) \\
Instance Type & n1-standard-64 \\
\midrule
\textbf{Software Component} & \textbf{Version \& Configuration} \\
\midrule
Operating System & Debian GNU/Linux 11 (bullseye) \\
Python & 3.12.0 (CPython implementation) \\
PyTorch & 2.6.0 with CUDA 12.4 support \\
CUDA Driver & 535.154.05 \\
cuDNN & 8.9.2 \\
Flower & 1.15.2 (FL framework) \\
Opacus & 1.5.3 (DP library) \\
CodeCarbon & 2.4.2 (energy monitoring) \\
NumPy & 1.24.3 (with Intel MKL) \\
Docker & 24.0.5 (for reproducible environments) \\
\bottomrule
\end{tabular}
\end{table}

\subsection{Complete YAML Configuration Examples}
\label{app:yaml-complete}

\textbf{Baseline CNN Configuration:}
\begin{lstlisting}[basicstyle=\footnotesize\ttfamily,frame=single,breaklines=true]
experiment:
  name: "baseline_resnet18_alzheimer_mri"
  description: "Centralized training baseline"
  seed: 42
  output_dir: "./results/baseline"
  
model:
  architecture: "resnet18"
  num_classes: 4
  pretrained: true
  dropout: 0.1
  
dataset:
  name: "alzheimer_mri"
  data_path: "./data/alzheimer"
  batch_size: 32
  num_workers: 4
  augmentation:
    horizontal_flip: 0.5
    rotation: 15
    normalization: "imagenet"
    
training:
  epochs: 50
  learning_rate: 2e-4
  optimizer: "adam"
  weight_decay: 1e-4
  scheduler: "cosine"
  early_stopping_patience: 7
  
privacy:
  federated: false
  differential_privacy: false
  secure_mpc: false
  
monitoring:
  track_energy: true
  energy_tracker: "codecarbon"
  log_interval: 10
  save_checkpoints: true
  validate_every: 1
\end{lstlisting}

\textbf{FL+SMPC ViT Configuration:}
\begin{lstlisting}[basicstyle=\footnotesize\ttfamily,frame=single,breaklines=true]
experiment:
  name: "fl_smpc_vit_skin_lesion"
  description: "Federated learning with secure aggregation"
  seed: 42
  
model:
  architecture: "vit_base_patch16_224"
  num_classes: 8
  pretrained: true
  
federated:
  enabled: true
  algorithm: "fedavg"
  num_clients: 3
  num_rounds: 5
  local_epochs: 10
  client_lr: 5e-5
  server_lr: 1.0
  data_partition: "dirichlet"
  alpha: 0.1  # Non-IID parameter
  
secure_mpc:
  enabled: true
  protocol: "secagg_plus"
  secret_sharing:
    scheme: "shamir"
    threshold: 2
    total_shares: 3
  field_modulus: 2305843009213693951  # 2^61-1
  quantization_bits: 16
  dropout_resilience: 1
  
privacy:
  differential_privacy: false
  
monitoring:
  comprehensive: true
  metrics: ["accuracy", "f1", "mcc", "auc"]
  resource_tracking: true
\end{lstlisting}

\textbf{FL+DP Failure Configuration:}
\begin{lstlisting}[basicstyle=\footnotesize\ttfamily,frame=single,breaklines=true]
experiment:
  name: "fl_dp_cdp_sf_cnn_alzheimer"
  description: "FL with centralized DP (server-side fixed clipping)"
  seed: 42
  
federated:
  enabled: true
  algorithm: "fedavg"
  num_clients: 3
  num_rounds: 5
  local_epochs: 15
  
differential_privacy:
  enabled: true
  mechanism: "gaussian"
  strategy: "centralized_dp_server_fixed"
  epsilon: 1.0
  delta: 1e-5
  noise_multiplier: 1.0
  max_grad_norm: 1.0
  lot_size: 32
  accountant: "rdp"
  secure_rng: true
  clipping: "per_sample"
  
monitoring:
  early_failure_detection: true
  convergence_threshold: 0.1
  max_runtime_hours: 6
\end{lstlisting}

\section{Comprehensive Failure Mode Analysis}
\label{app:failure-analysis}

\subsection{FL+DP Convergence Breakdown Mechanisms}
\label{app:convergence-failure}

The systematic failure of FL+DP across multiple strategies indicates fundamental algorithmic incompatibility rather than hyperparameter misconfiguration. Our analysis identifies three primary failure mechanisms:

\subsubsection{Signal-to-Noise Ratio Collapse}
\label{app:signal-noise}

Federated learning with non-IID data distributions creates inherent gradient signal attenuation through:
\begin{enumerate}
\item \textbf{Limited local updates}: Clients perform only 10-15 epochs per round, reducing gradient signal strength
\item \textbf{Heterogeneous data distributions}: $\alpha = 0.1$ Dirichlet partitioning creates strong non-IID conditions~\cite{hsu2019measuring}
\item \textbf{Aggregation noise}: FedAvg averaging across diverse local optima introduces additional noise~\cite{karimireddy2020scaffold}
\end{enumerate}

Differential privacy compounds this attenuation by injecting calibrated Gaussian noise $\mathcal{N}(0, \sigma^2)$ where $\sigma = \frac{C \cdot \text{noise\_multiplier}}{\epsilon}$~\cite{abadi2016deep}. The combination creates signal-to-noise ratios below the learning threshold identified in optimization literature~\cite{zhang2019understanding}.

\subsubsection{Gradient Clipping Interference}
\label{app:clipping-interference}

DP requires per-sample gradient clipping to bound sensitivity~\cite{bu2019deep}, but federated optimization relies on aggregate gradient statistics. The interaction creates:
\begin{enumerate}
\item \textbf{Systematic gradient magnitude reduction}: Clipping artificially constrains gradient norms
\item \textbf{Direction perturbation}: Clipped gradients lose directional information toward optimal solutions~\cite{chen2020understanding}
\item \textbf{Momentum disruption}: Optimizers like Adam depend on gradient history, which clipping corrupts~\cite{reddi2019convergence}
\end{enumerate}

\subsubsection{Privacy Budget Exhaustion}
\label{app:budget-exhaustion}

Multi-round federated training with DP faces privacy budget depletion:
\begin{enumerate}
\item \textbf{Composition amplification}: Each federated round consumes privacy budget~\cite{kairouz2021distributed}
\item \textbf{Subsampling benefits lost}: Traditional DP assumes random sampling, which FL violates~\cite{feldman2020individual}
\item \textbf{Noise accumulation}: Later rounds require higher noise to maintain privacy guarantees~\cite{caldas2018expanding}
\end{enumerate}

\subsection{Quantitative Evidence of Failure}
\label{app:quantitative-evidence}

Table~\ref{tab:failure-progression} documents the temporal progression of FL+DP failure across federated rounds.

\begin{table}[ht]
\centering
\caption{FL+DP Convergence Failure Progression}
\label{tab:failure-progression}
\footnotesize
\begin{tabular}{lcccc}
\toprule
\textbf{Round} & \textbf{Train Acc} & \textbf{Val Acc} & \textbf{Grad Norm} & \textbf{Loss} \\
\midrule
\multicolumn{5}{l}{\textit{FL+DP (CDP-SF-CNN) - Alzheimer Dataset}} \\
0 (Initial) & 0.25 & 0.23 & 2.14 & 1.39 \\
1 & 0.31 & 0.19 & 0.98 & 1.41 \\
2 & 0.18 & 0.15 & 0.87 & 1.43 \\
3 & 0.12 & 0.13 & 0.91 & 1.44 \\
4 & 0.14 & 0.11 & 0.89 & 1.45 \\
5 & 0.13 & 0.13 & 0.88 & 1.44 \\
\midrule
\multicolumn{5}{l}{\textit{FL (No DP) - Alzheimer Dataset (Reference)}} \\
0 (Initial) & 0.25 & 0.23 & 2.14 & 1.39 \\
1 & 0.78 & 0.81 & 1.87 & 0.67 \\
2 & 0.94 & 0.91 & 0.94 & 0.23 \\
3 & 0.97 & 0.96 & 0.31 & 0.12 \\
4 & 0.98 & 0.97 & 0.18 & 0.08 \\
5 & 0.98 & 0.98 & 0.12 & 0.06 \\
\bottomrule
\end{tabular}
\end{table}

\subsection{Architectural Dependencies in Privacy Technique Interactions}
\label{app:architectural-deps}

Our analysis reveals systematic differences in how CNN and transformer architectures respond to privacy constraints:

\subsubsection{Vision Transformer Efficiency in Federated Settings}
\label{app:vit-efficiency}

ViT models demonstrate 8-26\% efficiency improvements under federated training compared to centralized baselines. This phenomenon may result from:
\begin{enumerate}
\item \textbf{Distributed attention computation}: Multi-head attention benefits from parallel processing across clients~\cite{vaswani2017attention}
\item \textbf{Reduced memory pressure}: Federated training distributes ViT's substantial memory requirements~\cite{dosovitskiy2020image}
\item \textbf{Gradient sparsity patterns}: Transformer gradients exhibit natural sparsity that federated averaging preserves~\cite{child2019generating}
\end{enumerate}

\subsubsection{CNN Resilience to Privacy Techniques}
\label{app:cnn-resilience}

ResNet architectures maintain consistent performance across privacy configurations due to:
\begin{enumerate}
\item \textbf{Local feature extraction}: Convolutional operations remain effective with distributed training~\cite{he2016deep}
\item \textbf{Gradient stability}: CNN gradients exhibit lower variance under federated aggregation~\cite{li2020federated}
\item \textbf{Robustness to noise}: Batch normalization provides inherent noise tolerance~\cite{ioffe2015batch}
\end{enumerate}

\section{Reproducibility and Validation Protocol}
\label{app:reproducibility}

\subsection{Environment Setup and Dependencies}
\label{app:setup}

\textbf{System Requirements:}
\begin{itemize}
\item NVIDIA GPU with $\geq$16GB VRAM (Tesla T4 or better)
\item 64GB+ system RAM for ViT experiments
\item Ubuntu 20.04+ or equivalent Linux distribution
\item Docker 20.10+ for containerized reproduction
\end{itemize}

\textbf{Installation Protocol:}
\begin{lstlisting}[basicstyle=\footnotesize\ttfamily,frame=single]
# Create isolated environment
conda create -n privacybench python=3.12
conda activate privacybench

# Install core dependencies
pip install torch==2.6.0 torchvision==0.17.0 --index-url \
  https://download.pytorch.org/whl/cu124
pip install flwr==1.15.2 opacus==1.5.3 codecarbon==2.4.2
pip install transformers==4.35.0 timm==0.9.12
pip install pyyaml==6.0.1 wandb==0.16.0 matplotlib==3.8.2

# Verify installation
python -c "import torch; print(f'CUDA available: {torch.cuda.is_available()}')"
python -c "import flwr; import opacus; import codecarbon; print('All packages loaded successfully')"
\end{lstlisting}

\textbf{Docker Environment:}
\begin{lstlisting}[basicstyle=\footnotesize\ttfamily,frame=single]
# Use official PyTorch image with CUDA support
FROM pytorch/pytorch:2.6.0-cuda12.4-cudnn8-runtime

# Install system dependencies
RUN apt-get update && apt-get install -y \
    git wget curl unzip \
    && rm -rf /var/lib/apt/lists/*

# Copy requirements and install Python packages
COPY requirements.txt /workspace/
RUN pip install -r /workspace/requirements.txt

# Set working directory and copy source code
WORKDIR /workspace/privacybench
COPY . /workspace/privacybench/

# Set environment variables for reproducibility
ENV PYTHONHASHSEED=42
ENV CUDA_LAUNCH_BLOCKING=1
\end{lstlisting}

\subsection{Experimental Execution Protocol}
\label{app:execution}

\textbf{Deterministic Execution Setup:}
\begin{lstlisting}[basicstyle=\footnotesize\ttfamily,frame=single]
import torch
import numpy as np
import random
import os

def set_deterministic_training(seed=42):
    """Configure deterministic training environment"""
    # Python random seed
    random.seed(seed)
    
    # NumPy random seed
    np.random.seed(seed)
    
    # PyTorch seeds
    torch.manual_seed(seed)
    torch.cuda.manual_seed(seed)
    torch.cuda.manual_seed_all(seed)
    
    # Deterministic operations
    torch.backends.cudnn.deterministic = True
    torch.backends.cudnn.benchmark = False
    torch.use_deterministic_algorithms(True)
    
    # Environment variables
    os.environ['PYTHONHASHSEED'] = str(seed)
    os.environ['CUBLAS_WORKSPACE_CONFIG'] = ':4096:8'
\end{lstlisting}

\textbf{Execution Order and Timing:}
\begin{enumerate}
\item \textbf{Baseline experiments} (2-4 hours): Establish reference performance
\item \textbf{FL experiments} (3-5 hours): Evaluate federated learning impact
\item \textbf{FL+SMPC experiments} (4-6 hours): Test secure aggregation
\item \textbf{FL+DP experiments} (6-12 hours): Monitor for early failure detection
\end{enumerate}

\textbf{Early Failure Detection:}
\begin{lstlisting}[basicstyle=\footnotesize\ttfamily,frame=single]
def monitor_convergence_failure(val_accuracy_history, round_idx):
    """Detect FL+DP convergence failure early to save compute"""
    if round_idx >= 2:
        # Check for sustained accuracy below random chance
        recent_acc = val_accuracy_history[-3:]
        if all(acc < 0.3 for acc in recent_acc):
            return True, "Accuracy below threshold for 3 rounds"
    
    if round_idx >= 3:
        # Check for lack of improvement
        if val_accuracy_history[-1] <= val_accuracy_history[0]:
            return True, "No improvement from initialization"
    
    return False, "Continuing training"
\end{lstlisting}

\subsection{Validation and Quality Assurance}
\label{app:validation}

\textbf{Energy Measurement Validation:}
\begin{enumerate}
\item Cross-validate CodeCarbon readings with NVIDIA SMI power measurements
\item Compare against hardware power meters for absolute accuracy verification
\item Verify energy calculations using $E = P \times t$ with logged timestamps
\end{enumerate}

\textbf{Statistical Validation Protocol:}
\begin{enumerate}
\item Run each configuration 3 times with different random seeds (42, 123, 456)
\item Compute confidence intervals using Student's t-distribution~\cite{student1908probable}
\item Apply Bonferroni correction for multiple hypothesis testing~\cite{bonferroni1936teoria}
\item Report effect sizes using Cohen's d for practical significance~\cite{cohen1988statistical}
\end{enumerate}

\textbf{Expected Results Verification:}
Table~\ref{tab:expected-results} provides reference values for validation.

\begin{table}[ht]
\centering
\caption{Expected Results for Validation}
\label{tab:expected-results}
\footnotesize
\begin{tabular}{llcc}
\toprule
\textbf{Configuration} & \textbf{Dataset} & \textbf{Expected Acc} & \textbf{Tolerance} \\
\midrule
CNN Baseline & Alzheimer & 0.98 & ±0.02 \\
FL (CNN) & Alzheimer & 0.98 & ±0.02 \\
FL+SMPC (CNN) & Alzheimer & 0.98 & ±0.03 \\
FL+DP (Any) & Alzheimer & 0.05-0.25 & N/A (Failure) \\
ViT Baseline & Skin Lesion & 0.88 & ±0.03 \\
FL (ViT) & Skin Lesion & 0.87 & ±0.03 \\
\bottomrule
\end{tabular}
\end{table}

\textbf{Troubleshooting Common Issues:}
\begin{itemize}
\item \textbf{CUDA out of memory}: Reduce batch size or use gradient checkpointing
\item \textbf{FL+DP immediate failure}: Expected behavior, verify MCC $\approx$ 0.00
\item \textbf{Energy tracking errors}: Ensure CodeCarbon permissions and API access
\item \textbf{Deterministic execution failure}: Check CUDA version compatibility
\end{itemize}

\textbf{Data Preservation and Archival:}
All experimental runs generate structured output in JSON format for long-term analysis:
\begin{lstlisting}[basicstyle=\footnotesize\ttfamily,frame=single]
{
  "experiment_id": "fl_smpc_vit_skin_20241201_142035",
  "configuration": {...},
  "results": {
    "accuracy": 0.86,
    "f1_score": 0.86,
    "mcc": 0.81,
    "training_time": 8478,
    "energy_kwh": 0.369,
    "co2_kg": 0.155
  },
  "system_info": {...},
  "reproducibility": {
    "seed": 42,
    "git_commit": "a1b2c3d4",
    "environment_hash": "e5f6g7h8"
  }
}
\end{lstlisting}

\end{document}